\documentclass[twocolumn,showpacs,preprintnumbers,amsmath,amssymb]{revtex4}

\usepackage{graphicx}
\usepackage{dcolumn}
\usepackage{bm}
\begin{document}

\preprint{APS/123-QED}

\title{Large magnetoresistance in $\pi$-conjugated semiconductor thin film devices}

\author{\"{O}. Mermer}
\address{Department of Physics and Astronomy, The University of Iowa, Iowa City, IA 52242-1479, USA}

\author{G. Veeraraghavan, T. L. Francis}
\address{Department of Electrical and Computer Engineering, The University of Iowa, Iowa City, IA 52242-1595, USA}

\author{Y. Sheng, D. T. Nguyen, M. Wohlgenannt}\email{markus-wohlgenannt@uiowa.edu}
\address{Department of Physics and Astronomy, The University of Iowa, Iowa City, IA 52242-1479, USA}

\author{A. K\"{o}hler}
\address{Institute of Physics, University of Potsdam, Am Neuen Palais 10, 14469 Potsdam, Germany}

\author{M. K. Al-Suti, M.S. Khan}
\address{ Department of Chemistry, Sultan Qaboos University, P. O. Box 36, Al Khod 123, Sultanate of Oman}

\date{\today}

\begin{abstract}
Following the recent discovery of large magnetoresistance at room temperature in polyfluorence sandwich devices, we have performed a comprehensive magnetoresistance study on a set of organic semiconductor sandwich devices made from different pi-conjugated polymers and small molecules. The measurements were performed at different temperatures, ranging from 10K to 300K, and at magnetic fields, $B < 100mT$. We observed large negative or positive magnetoresistance (up to 10$\%$ at 300K and 10mT) depending on material and device operating conditions. We compare the results obtained in devices made from different materials with the goal of providing a comprehensive picture of the experimental data. We discuss our results in the framework of known magnetoresistance mechanisms and find that none of the existing models can explain our results.
\end{abstract}

\pacs{Valid PACS appear here}

\maketitle
\section{\label{sec:level1}Introduction}

Organic $\pi$-conjugated materials have been used to manufacture devices such as organic light-emitting diodes (OLEDs) \cite{ElectroluminescenceReview}, photovoltaic cells \cite{PVCellReview} and field-effect transistors \cite{reviewFET}. Recently there has been a growing interest in spin \cite{nature,TalianiPaper,ValyNature} and magnetic field effects \cite{tomshortpaper,Mermer:2005,Preprint,Kalinowski:2003,Kalinowski:2004,Bussmann,Yoshida:2005,Salis:2004} in these materials. During the study of sandwich devices made from the $\pi$-conjugated polymer polyfluorene (PFO) we recently discovered a large and intriguing magnetoresistive (MR) effect, which we dubbed organic magnetoresitance (OMAR). OMAR may find application in magnetic field sensors, e.g. in OLED interactive displays (patent pending) \cite{tomshortpaper}. To the best of our knowledge, this work was the first to actually demonstrate room-temperature MR in organic semiconductors with a magnitude and signal to noise ratio sufficient for application. Here, we report a comprehensive MR study in both polymer and small molecular sandwich devices in order to complete a basic magnetotransport characterization of these systems. In the following we present a large body of experimental data we have collected to characterize the MR effect in different polymers, namely PFO, regio-regular Poly(3-hexylthiophene-2,5-diyl) (RR-P3HT), regio-random Poly(3-octylthiophene-2,5-diyl)(RRa-P3OT), Poly(5,8-diethynyl-2,3-diphenylquinoxaline) (organic PPE), Pt-containing Poly(5,8-diethynyl-2,3-diphenylquinoxaline) (Pt-containing PPE) as well as small molecules such as Tris-(8-hydroxyquinoline) aluminium (Alq3) and Pentacene.

\begin{figure}
\includegraphics[width=\columnwidth]{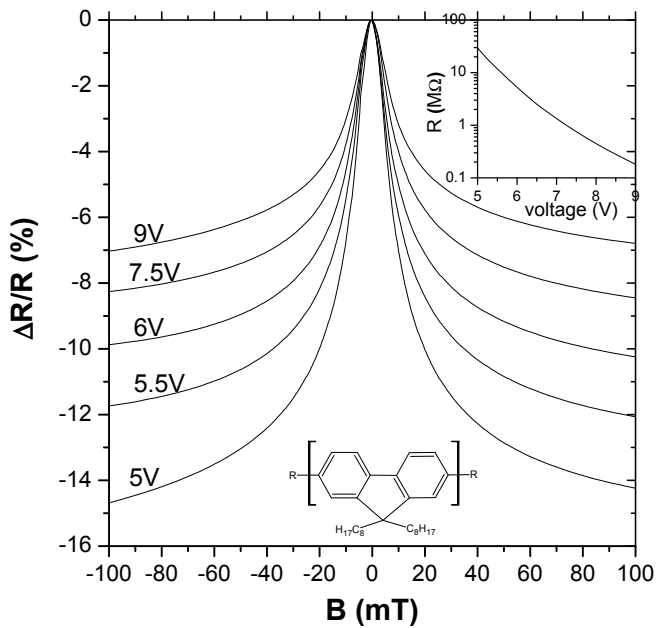}
\caption{\label{fig:PedotPfo} Magnetoresistance, $\Delta R/R$ curves, measured at room temperature in an ITO/PEDOT/PFO ($\approx$ 100 nm)/Ca device at different voltages. The inset shows the device resistance as a function of the applied voltage.}
\end{figure}

\section{Experimental}

PFO was purchased from American Dye Source, inc.. Alq3 was purchased from H. W. Sands Corp. RR-P3HT, RRa-P3OT and Pentacene were purchased from Sigma Aldrich. The 5,8-diethynyl-2,3-diphenylquinoxaline unit and its Pt-containing polymer were prepared according to published procedures \cite{Khan,Wilson}. The organic polymer was synthesized by palladium-catalyzed polycondensation of 1,4-bis(n-octyloxy)-2,5-diiodobenzene and 5,8-diethynyl-2,3-diphenylquinoxaline in a 1:1 ratio. The fabrication of the organic sandwich devices started with glass substrates coated with 40nm of ITO, purchased from Delta Technologies. The conducting polymer Poly (3,4-ethylenedioxythiophene)-poly (styrenesulfonate) (PEDOT:PSS), purchased from H. C. Starck was spin coated at 2000 rpm on top of the ITO in some devices to provide an efficient hole injecting electrode. All other manufacturing steps were carried out in a nitrogen glove box. The active polymer film was spin coated onto the substrate from a chloroform solution. The semiconductor film thickness was varied by using different concentrations of polymer solution (5 - 30 mg/ml). The small molecular film layers were made by thermal evaporation. A Ca cathode followed by capping layer of Al were then deposited by thermal and electron beam evaporation, respectively at a base pressure less than 1$\times10^{-6}$ mbar on top of the organic thin films. The general device structure used for our measurements was metal/organic semiconductor/metal. We studied both soxhlet purified RR-P3HT and as received RR-P3HT. The initial soxhlet purification step was performed for 8-12 hours using hexane to remove low molecular weight components. The second step was performed for 24 hours using methanol to remove metallic impurities. The third step was performed for 8-12 hours using chloroform to dissolve the pure polymer. A roto-vap was used to separate the polymer from the solution.

The samples were mounted on the cold finger of a closed-cycle He cryostat located between the poles of the electromagnet. MR was determined by measuring the current at a constant applied voltage for different magnetic fields, B.

\begin{figure}
\includegraphics[width=\columnwidth]{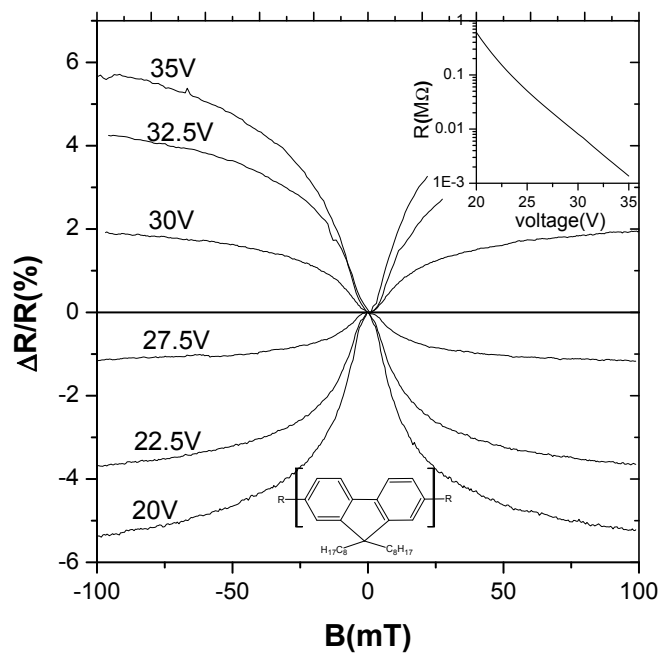}
\caption{\label{fig:Pfo200K} Magnetoresistance, $\Delta R/R$ curves,
measured at 200K in an ITO/PFO ($\approx$ 60
nm)/Ca device at different voltages. The inset shows the device
resistance as a function of the applied voltage.}
\end{figure}

\section{ Experimental Results }

\subsection{\label{sec:level1} Organic Magnetoresistance in polymer devices}
\subsubsection{\label{sec:level1} Polyfluorene Devices}

We originally discovered OMAR in PFO sandwich devices \cite{tomshortpaper}. OMAR in PFO is treated in detail in the original publication \cite{tomshortpaper}, here we show only the room temperature data. This data is shown in Fig.~\ref{fig:PedotPfo} for different applied voltages. We summarize the findings of our previous work: $\bullet$ We observed that OMAR curves are independent of the angle between film plane and the applied magnetic field and also the sign of the magnetic field. $\bullet$ We found that the measured OMAR effect does not critically depend on the choice of the top electrode (cathode) materials. This indicates that the observed OMAR effect is due to hole transport since it occurs also in hole only devices, namely those with Au cathodes where only very weak electroluminescence is observed. $\bullet$ We fabricated devices with different polymer film thickness and found that the observed OMAR effect was similar in all devices independent of PFO film thickness (except for an increase in turn on voltage). This clearly suggests that this MR effect is a bulk effect rather than an interface (electrode) effect. $\bullet$ We found that OMAR in PFO strongly depends on the choice of anode material. In particular, using PEDOT as the anode results  in a significant reduction in the onset voltage and an increase in the observed  OMAR effect. The reduced onset voltage and increased MR can be rationalized considering the decrease in the hole-injection barrier and the resulting reduction of the interface resistance. $\bullet$ In Fig.~\ref{fig:PedotPfo} it can be seen that $\Delta R/R$ typically increases in magnitude with  increasing R. $\bullet$ We found that the OMAR effect is not related to (unintentional/intentional) impurities, such as left over catalysts from the polymerization reaction. Elemental analysis (performed by ADS) of our ADS PFO showed signals only for Ni impurities at levels less than 20 ppm. In addition to that, we also tried several batches of PFO with different Ni contents (21, 177, 683, 3460 and 8840 ppm) and all measurements showed that there is no significant dependence of the MR effect on the Ni impurity concentration.

In Fig.~\ref{fig:Pfo200K} we also show measured OMAR curves in an ITO/PFO($\approx 60nm$)/Ca device. We found that the ITO/PFO interface is less suitable for hole injection than the PEDOT/PFO interface; resulting in a large increase in turn-on voltage of the device. The measurements have been performed at 200K to reduce thermal drift. Importantly, we also observed positive MR \cite{Preprint} in the ITO anode device at high applied voltages. This shows that OMAR may be either negative or positive, dependent on device parameters and operating conditions.

\subsubsection{\label{sec:level1} Regio-Regular and Regio-Random Polythiophene Devices}

Next we extend our study to include RR-P3HT and RRa-P3OT. In contrast to PFO, RR-P3HT and RRa-P3OT do not contain benzene rings. RR-P3HT was selected due to its high mobility among polymers and its well known usage in transistors, smart pixels and optoelectronic devices \cite{Sirringhaus,Sirringhaus2,scienceVardeny}. The reason for the high carrier mobilities in transistors is the self-organization of RR-P3HT chains resulting in a lamellae structure perpendicular to the film substrate \cite{Sirringhaus,vardenyRR}. Delocalization of the charge carriers among the lamellae has been invoked to be the reason for the high interlayer mobility, with values reported as high as  0.1 cm$^2$V$^{-1}$s$^{-1}$ \cite{Sirringhaus}. RRa-P3OT, which has a similar chemical structure as RR-P3HT, has a lamellae structure to a lesser degree and the carrier mobility is consequently much smaller, around 10$^{-4}$ cm$^2$V$^{-1}$s$^{-1}$ \cite{Sirringhaus,vardenyRR}.

\begin{figure*}
\includegraphics[width=1.5\columnwidth]{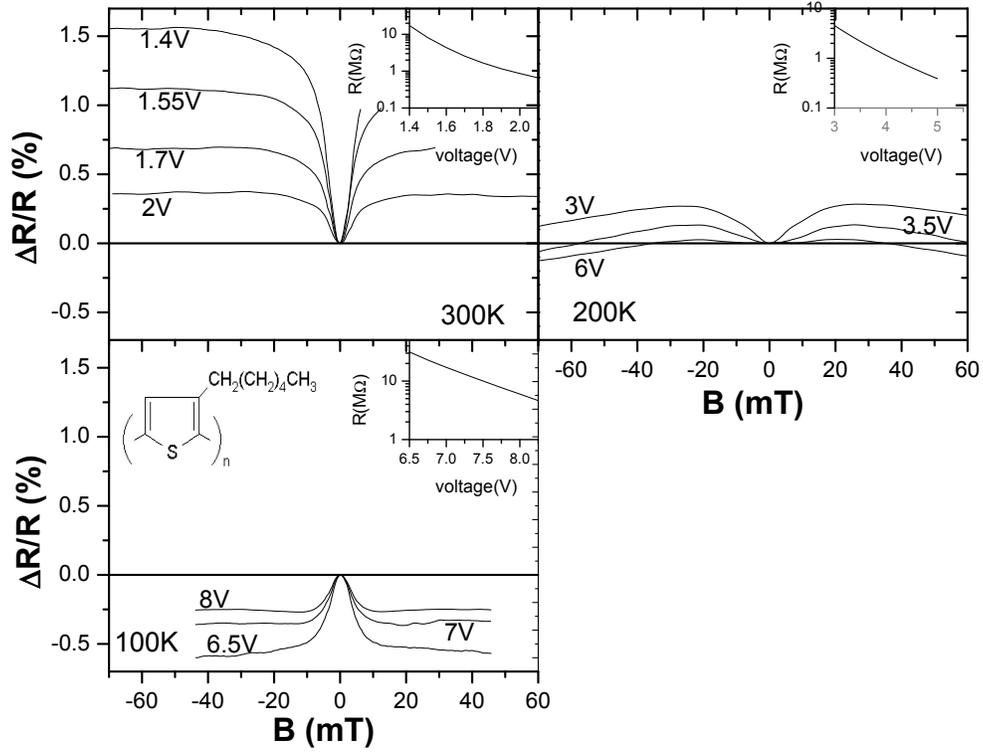}
\caption{\label{fig:regular} Magnetoresistance, $\Delta R/R$ curves (both positive and negative) in an ITO/PEDOT/RR-P3HT($\approx100nm$)/Ca device measured at different temperatures (100K, 200K, and 300K). The insets show the device resistance as a function of the applied voltage.}
\end{figure*}

Fig.~\ref{fig:regular} shows measured OMAR traces as a function of temperature in an ITO/PEDOT/RR-P3HT/Ca device.
The magnitude of the observed OMAR effect is small (less than 1.5 $\%$). The data shows that the OMAR effect can be both positive and negative in RR-P3HT, mostly dependent on temperature. At room temperature, the OMAR effect is completely positive, whereas at 100K the effect is negative. At 200K, a transition from positive to negative MR occurs as the voltage increases in RR-P3HT. A similar transition occurs in PFO when the voltage is decreased. This intriguing behavior may hold a clue for identifying the mechanism responsible for the MR effect. We studied both soxhlet purified RR-P3HT and as received RR-P3HT. We did not find any significant difference between the two samples. This supports our conclusion from the experiments in PFO devices that OMAR is not caused by impurities.

We observed only negative OMAR traces in ITO/PEDOT/RRa-P3OT/Ca devices, shown in Fig.~\ref{fig:random}, at all temperatures spanning the range between 10K and 300K. The data measured in RRa-P3OT is noisier than in RR-P3HT, presumably because of the increased disorder and lower purity of the RRa-P3OT polymer. The magnitude of the measured OMAR effect was found to be significantly higher than that of the RR-P3HT device. Therefore increasing disorder appears to enhance the OMAR effect.

\begin{figure*}
\includegraphics[width=1.5\columnwidth]{TempRandom.prn}
\caption{\label{fig:random} Magnetoresistance, $\Delta R/R$ curves in an ITO/PEDOT/RRa-P3OT($\approx100nm$)/Ca device measured at different temperatures (10K, 100K, 200K, and 300K). The insets show the device resistance as a function of the applied voltage.}
\end{figure*}

\subsubsection{\label{sec:level1} Polyphenylene ethynelene (PPE) and Pt-containing PPE: the influence of spin-orbit coupling}

\begin{figure*}
\includegraphics[width=1.5\columnwidth]{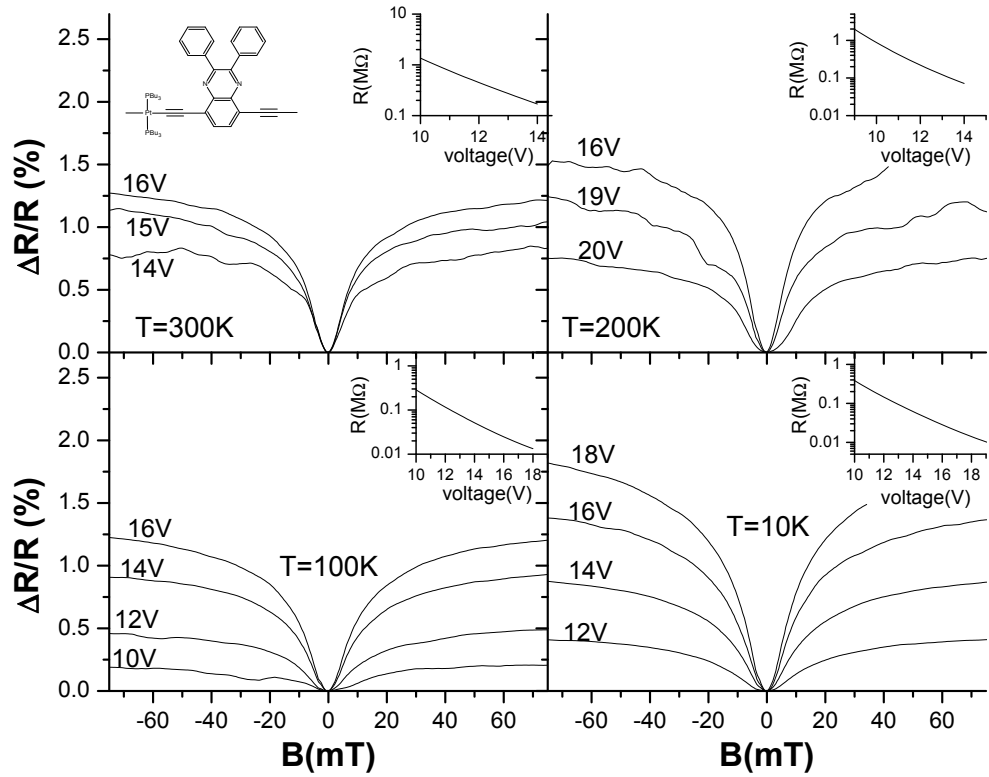}
\caption{\label{fig:P42} Magnetoresistance, $\Delta R/R$ curves in an ITO/PEDOT/Pt-PPE/Ca device measured at different temperatures (10K, 100K, 200K, and 300K). The insets show the device resistance as a function of the applied voltage.}
\end{figure*}

\begin{figure*}
\includegraphics[width=1.5\columnwidth]{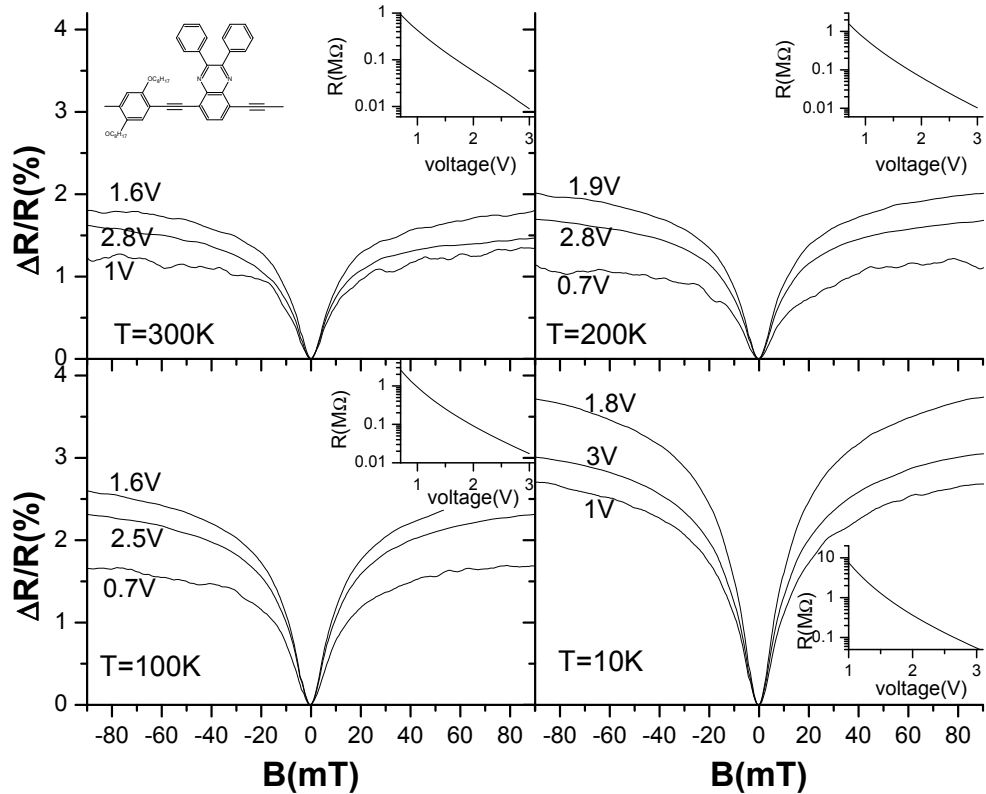}
\caption{\label{fig:P49} Magnetoresistance, $\Delta R/R$ curves in an ITO/PEDOT/PPE/Ca device measured at different temperatures (10K, 100K, 200K, and 300K). The insets show the device resistance as a function of the applied voltage.}
\end{figure*}

We studied OMAR in two pi-conjugated polymers containing triple bonds, namely i) the Pt-containing polymer shown in Fig.~\ref{fig:P42}, and ii) a very similar polymer without Pt shown in Fig.~\ref{fig:P49}. The heavy Pt atom introduces relatively strong spin-orbit coupling \cite{wilsonnature} and the spin orientation ceases to be a good quantum number. It was shown previously that the Pt atom does not interrupt the conjugation \cite{wilsonnature,vardenyPRB}. We studied these polymers to examine a possible interdependence between spin-orbit coupling and the OMAR effect.

Figs.~\ref{fig:P42} and \ref{fig:P49} show OMAR traces measured in ITO/PEDOT/Pt-PPE/Ca and ITO/PEDOT/PPE/Ca devices, respectively, measured at different temperatures. It is seen that the OMAR effect is positive and is quite similar at all temperatures for both polymers. We therefore obtain the result that spin-orbit coupling has apparently little effect on OMAR. This in turn suggests that OMAR is not related to spin but is likely related to an orbital effect. However, more future work, e.g. on oriented polymer films, is necessary to confirm this conclusion.

\subsection{\label{sec:level1} Organic Magnetoresistance in small molecule devices}

\begin{figure}
\includegraphics[width=\columnwidth]{MR_Alq3_RT.prn}
\caption{\label{fig:Alq3RT} Magnetoresistance, $\Delta R/R$ curves, measured at room temperature
in an ITO/PEDOT/Alq$_3$($\approx100nm$)/Ca device at different voltages. The insets show the
device resistance as a function of the applied voltage.}
\end{figure}

\begin{figure*}
\includegraphics[width=1.5\columnwidth]{TempPentacene.prn}
\caption{\label{fig:pentacene} Magnetoresistance, $\Delta R/R$ curves (both positive and negative) in an ITO/PEDOT/Pentacene($\approx100nm$)/Ca device measured at different temperatures (10K, 100K, 200K, and 300K). The insets show the device resistance as a function of the applied voltage.}
\end{figure*}

Having demonstrated OMAR in polymers, it is natural to ask whether OMAR also exists in small molecules, e.g. the prototypical small molecule Alq$_3$. This extension would be highly relevant both from an application as well as a scientific point of view. Whereas polymers are quasi-one-dimensional, Alq$_3$ corresponds more to quasi-zero-dimensional. Whereas PFO and most other $\pi$-conjugated polymers are hole-conductors, meaning that the hole mobility greatly exceeds that for electrons \cite{Redecker:1998}, Alq$_3$ is an electron transporter \cite{Kepler:1995}. In addition, in polymers it was found that the interaction cross sections between electrons and holes are spin-dependent \cite{nature,wilsonnature}, whereas they were found to be spin-independent in Alq$_3$ \cite{Baldo,rAlq3}. Therefore it is non-trivial that OMAR would occur in Alq$_3$ even if it occurs in polymers.

A detailed study of the OMAR effect in Alq$_3$ devices can be found in the original publication \cite{Mermer:2005}. Here only the room temperature data are shown in Fig.~\ref{fig:Alq3RT}. It can be seen that a large OMAR effect can also be achieved in small molecules. After a summary of our previous results in Alq$_3$, we will extend our study in small molecules to include pentacene. $\bullet$ As in PFO, the effect is independent of the sign and direction of the magnetic field, and shows only a weak temperature dependence. $\bullet$ In distinction to our results in polymers, we found that both I-V and OMAR responses critically depend on the choice of electron-injecting cathode material. Ca cathodes result in low onset voltage and large OMAR response, whereas using Al results in a drastic increase in onset V and decrease in OMAR magnitude at small currents. The situation is even more drastic when using a Au cathode. The increased onset voltage and decreased OMAR can be rationalized considering the increase in the electron-injection barrier and the resulting increase of the interface resistance, respectively, when using high work function cathodes (Ca has the lowest work-function, followed by Al, whereas Au has one of the largest work functions). This strong dependence of OMAR in Alq$_3$ on the choice of cathode material was to be expected since the hole mobility is about 100 times smaller than the electron mobility \cite{Kepler:1995} in Alq$_3$. $\bullet$ In Ref. \cite{Mermer:2005} we also examined the magnetic field effect on the electroluminescence and electroluminescence efficiency. We found that most of the magnetic field effect (MFE) on the electroluminescence is simply a result of the MR effect, since a change in current trivially implies a change in electroluminescence intensity. However, even when running the device at a constant current, a small but non-zero MFE remains that may be attributable to an MFE on the emission efficiency of the device. Here we note that all the prior work on MFE \cite{Kalinowski:2003,Kalinowski:2004,Bussmann,Yoshida:2005} we know of assumed that the magnetic field affects the emission efficiency directly, whereas the MR effect was largely neglected.

Fig.~\ref{fig:pentacene} shows the temperature dependence of OMAR in an ITO/PEDOT/Pentacene/Ca device. OMAR in the pentacene device is much smaller compared to the Alq$_3$ device. The picture of OMAR in pentacene is relatively complex. We found that OMAR can be positive or negative at 10K, 200K and 300K. At these temperatures the transition from positive to negative MR occurs as the bias voltage increases. OMAR at 100K follows however a different trend. The effect was negative at all biases studied, with the smallest bias giving the largest negative effect.

\begin{figure}
\includegraphics[width=\columnwidth]{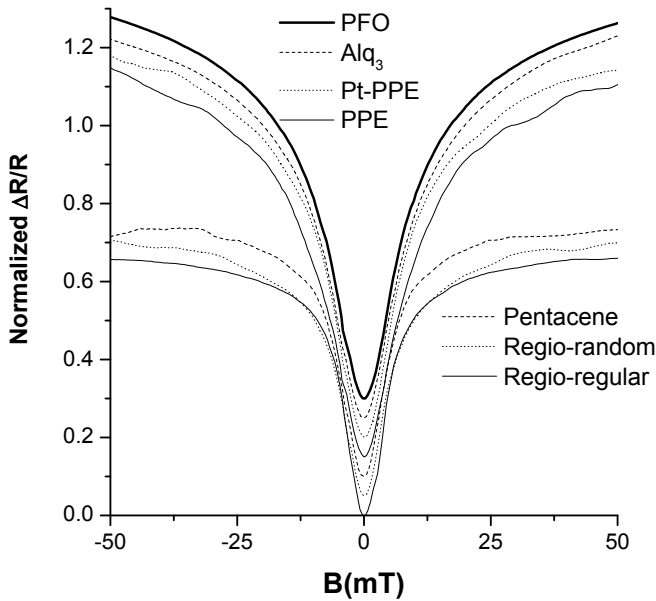}
\caption{\label{fig:universality} Normalized magnetoresistance, $\Delta R/R$ curve of PFO,
RRa-P3OT, RR-P3HT, Pt-PPE, PPE, Alq3, pentacene devices measured at room temperature. The
data have been offset for clarity}
\end{figure}

\section{\label{sec:level1} Discussion}

\subsection{\label{sec:level1} Universality of the OMAR effect}

Fig.~\ref{fig:universality} shows normalized $\Delta R/R$ traces of all the tested devices at room temperature. The normalization was done with respect to the magnitude of the OMAR effect at 50mT. It is seen from the figure that there are two groups of OMAR traces in organic semiconductor materials. One (pentacene, RR-P3HT and RRa-P3OT) where the OMAR effect has saturated at 50mT and the other (PFO, Alq3, Pt-PPE and PPE) where the effect is still unsaturated. The functional dependence of the OMAR effect was found to be similar in devices within a group. This is very surprising, since the chemical structure of these materials are quite different and therefore one expects them to possess varying materials parameters such as transport properties. The "universality" of the OMAR effect therefore implies that the explanation for the origin of OMAR must be quite general and simple.

\subsection{\label{sec:level1} Discussion of possible mechanisms}

We discuss possible mechanisms that could explain the observed OMAR effect. To the best of our knowledge, the mechanism causing OMAR is currently not known. Most MR mechanisms rely on the presence of ferromagnetic materials and are therefore not applicable to our devices. We are familiar with the following mechanisms
that cause MR in nonmagnetic materials: (1) Classical magnetoresistance, (2) hopping magnetoresistance
\cite{EfrosBook}, (3) electron-electron (e-e)interaction \cite{MRBook}, and (4) weak localization \cite{WeakLocalizationMetals}. Classical magnetoresistance (due to Lorenz force), results in positive MR due to the fact that the applied magnetic field causes the  electrons to go around circular orbits with the cyclotron frequency and hence the orbital motion increases resistance. This means that classical MR is  always a positive effect with a magnitude on the order of $\mu^2B^2$. Using a typical value for the mobility, $\mu \approx 10^{-4}$, we estimate $\Delta R/R \approx 10^{-20}$ at  $B \approx 10 mT$ for classical MR. The classical MR is therefore much too small to explain OMAR at similar fields. In hopping magnetoresistance, an applied magnetic field shrinks the electron wave function and this reduces the overlap between hopping sites leading to an increase in resistance of the system resulting in a positive MR effect. The size of this effect is only appreciable if the magnetic length, $\lambda$ is comparable to the hopping distance. In our case, $\lambda$ ($= \sqrt{\hbar/eB}$, where $\hbar$ and $e$ have their usual meaning) is around 200nm at 10mT which is much bigger than the hopping distance, which we estimate to be $\approx 10nm$. At lower temperatures or in highly disordered system, corrections to the transport due to e-e interaction become important. This is mainly due to the fact that carriers interact often when they diffuse slowly or when the system is highly disordered. It can be shown that the e-e interaction is modified in the presence of the magnetic field, however this occurs only if the thermal energy, kT, is less than or comparable to the Zeeman energy, $\Delta E = g\mu_{B}B$ where g is the g-factor and $\mu_{B}$ is the bohr magneton. In our case, g is approximately 2 and $\Delta E$ is around 1$\mu eV$ at a field of 10mT and therefore much smaller than kT in our experiments. Therefore this mechanism also fails to explain OMAR. Weak localization (WL) due to quantum corrections to the Drude-like transport is another mechanism for MR. This mechanism is very well known from the study of diffusive transport in metals and semiconductors \cite{WeakLocalizationMetals,WeakLocalizationFET,WLQuantumDot}. It is based on back scattering processes due to constructive quantum interference. When a magnetic field is applied to the system, the quantum interference is destroyed by the magnetic field if the phase-delay due to the enclosed magnetic flux exceeds the coherence length. Therefore the resistivity is decreased (negative MR effect). In WL theory, it is assumed that the spin-orbit coupling is weak. A strong spin-orbit coupling would cause weak anti-localization (WAL) leading to a positive MR effect due to destructive quantum interference.

Mechanisms (1) to (3) lead to only positive MR, whereas OMAR can also be negative. This implies that the observed MR should not be due to either (1),(2) or (3). Interestingly we note that the observed OMAR traces closely resemble MR traces due to weak localization (WL, negative MR) and weak antilocalization (WAL, positive MR). This might tempt us to analyze the MR data using the theory of weak localization. Such an attempt, however produces several surprising results that casts some doubt on this interpretation. For example, the weak temperature dependence of the effect is contrary to most if not all of the literature on WL and WAL in inorganic conductors. The interested reader may refer to our earlier work on the Los Alamos preprint server \cite{Preprint}.

Frankevich and co workers \cite{Frankevich:1992,Frankevich:1994} have shown that lifetimes of pairs of paramagnetic species (such as electrons and holes), which may be in singlet and triplet states, are very sensitive to external magnetic fields within the range of hyperfine interaction. In general, models involving pairs of electrons and holes appear promising since typical magnetic dipole fields between pairs of electrons and holes can be expected to be on the order of $10 mT$. However, it is not quite clear how the pairing mechanism should occur in hole-only devices, such as the Au-cathode PFO devices we studied. One might also expect that this mechanism should depend strongly on the carrier density, whereas OMAR is only weakly dependent on current density. In particular, intersystem crossing between singlet and triplet electron-hole pairs resulting from hyperfine interaction have often been employed to explain the MFE on the electroluminescence in organic devices \cite{Kalinowski:2003,Yoshida:2005}. In this model, the external magnetic field results in Zeeman splitting. If the Zeeman splitting exceeds the hyperfine interaction strength (this typically occurs at $B \approx 1 mT$), then mixing between singlets and the two extremal Zeeman levels is no longer possible, therefore the intersystem crossing rate is reduced. However, our observation that OMAR traces are unchanged in Pt-containing PPE compared to those in PPE, implies that this model cannot account for OMAR. This is because the heavy Pt atom leads to an intersystem crossing rate that is very much stronger than that due to hyperfine interaction. Therefore much larger B should be required to prevent mixing between singlet and triplet levels as a result of the Pt-atom.

It therefore appears that a novel explanation for the observed MR effect needs to be found. This could lead to a better understanding of the transport processes in organic semiconductors. Follow-up experiments performed on current-in-plane devices, and in devices using crystalline or oriented organic semiconductors will likely provide further clues to the origin of the OMAR effect.

\section{ Conclusion}
In summary, we studied a recently discovered MR effect in various different pi-conjugated polymer and small molecular thin film devices. In PFO devices we found that the magnitude of the effect is $\approx$ 10\% at fields on the order of 10mT and can be either positive or negative, depending on operating conditions. The effect is also independent of the sign and direction of the magnetic field, and is only weakly temperature dependent. The OMAR effect appears to be a bulk effect related to the hole current. We also studied RR-P3HT and RRa-P3OT $\pi$-conjugated polymers to study the effect of order/disorder on OMAR. We found that the magnitude of the measured OMAR effect of RRa-P3OT device was higher than that of the RR-P3HT device. This might suggest that increasing disorder appears to enhance the OMAR effect. We also studied PPE and Pt-containing PPE polymers to check for a possible interrelation between spin-orbit coupling and the OMAR effect. We found that spin-orbit coupling has apparently little effect on OMAR. We also extended our study to small molecular devices. We observed a large OMAR effect in Alq$_3$ devices that is similar in size to that in the PFO devices. As in polymers, the effect is independent of the sign and direction of the magnetic field, and is only weakly temperature dependent. However, the OMAR effect in Alq$_3$ appears to be dominated by the electron current. For pentacene devices, we observed both negative and positive OMAR effect, whereas we observed only a negative effect in the Alq$_3$ devices. Strikingly, we found that the functional dependence of the OMAR effect on B was very similar in all devices we studied. The universality of the OMAR effect therefore implies that the explanation for the mechanism must be quite general and simple. To the best of our knowledge, this effect is not adequately described by any of the MR mechanisms known to date. It may therefore be related to the peculiar mode of charge transport in organic semiconductors, a field that still relatively little is known about. If one could find an explanation for this effect, this may lead to a breakthrough in the scientific understanding of organic semiconductors.

We acknowledge fruitful and inspiring discussions with Profs. M.
E. Flatt\'{e} and Z. V. Vardeny. This work was supported by University of Iowa Carver grant and NSF ECS 04-23911.

\bibliography{PRBpaper}
\end{document}